\documentclass{emulateapj}

\shorttitle{Gas Shepherding}
\shortauthors{Chang}

\begin{document}

\title{Gas Shepherding by an Infalling Satellite} 

\author{Philip Chang\altaffilmark{1,2}} 

\altaffiltext{1} {Astronomy Department and Theoretical Astrophysics
  Center, 601 Campbell Hall, University of California, Berkeley, CA
  94720; pchang@astro.berkeley.edu} 

\altaffiltext{2} {Miller Institute for Basic Research}

\begin{abstract}
  I calculate the action of a satellite, infalling through dynamical
  friction, on a coplanar gaseous disk of finite radial extent.  The
  disk tides, raised by the infalling satellite, couple the satellite
  and disk.  Dynamical friction acting on the satellite then shrinks
  the radius of the coupled satellite-disk system.  Thus, the gas is
  ``shepherded'' to smaller radii.  In addition, gas shepherding
  produces a large surface density enhancement at the disk edge.  If
  the disk edge then becomes gravitationally unstable and fragments,
  it may give rise to enhanced star formation. On the other hand, if
  the satellite is sufficiently massive and dense, the gas may be
  transported from $\sim 100$ pc to inside of a 10 to 10s of parsecs
  before completely fragmenting into stars.  I argue that gas
  shepherding may drive the fueling of active galaxies and central
  starbursts and I compare this scenario to competing scenarios.  I
  argue that sufficiently large and dense super star clusters (acting
  as the shepherding satellites) can shepherd a gas disk down to ten
  to tens of parsecs.  Inside of ten to tens of parsecs, another
  mechanism may operate, i.e., cloud-cloud collisions or a marginally
  (gravitationally) stable disk, that drives the gas $\lesssim 1$ pc,
  where it can be viscously accreted, feeding a central engine.
\end{abstract}


\section{Introduction}\label{sec:intro}

The tidal interaction between small satellites and the gas or particle
disks in which they are embedded is a subject of wide study in the
planetary community.  These satellites are tidally coupled via disk
tides, i.e., satellites excite spiral density waves at the Lindblad
resonances in the disk (Goldreich \& Tremaine 1978, 1980; Artymowicz
1993).  As a result of these tidal interactions, gaps can be opened up
in the embedded disks as in the case of the shepherding moons of
Saturn's rings (Goldreich \& Tremaine 1978) or in type-II migration
(Ward 1997).  Alternatively, if a gap does not open up, satellite may
migrate inward rapidly due to tidal torque imbalances, i.e., type-I
migration (Ward 1997).

The wide applicability of satellite-disk interactions in the planetary
community raises an interesting question of whether the same physics
may be applicable at larger scales, i.e., galactic scales.  In this
paper, I will address this question by considering a very simple
model, which illustrate the modification of the physics of
satellite-disk interaction when applied on a galactic scale.  Most
notably, the critical difference is that in addition to tidal torques,
the satellite will experience dynamical friction.  The inclusion of
dynamical friction produces a non-trivial effect.  Namely, dynamical
friction on the satellite provides a sink of angular momentum in the
system.  As a result, the coupled satellite-disk system will
continually lose angular momentum and sink toward the center.

To study this physics, I consider a very simple model.  In my model, a
satellite starts out in a circular orbit at a large radius and sinks
toward the central mass concentration because of dynamical friction on
the background stars.  Along its infall, it encounters a coplanar
gaseous disk or ring, which initially has a finite radial extent,
$r_{\rm d,0}$.  Tides begin to couple the satellite with the disk.
Because the satellite-disk system continues to suffer an ongoing loss
of angular momentum from dynamical friction, it will shrink in radius.
As a result gas can be transported on the dynamical friction timescale
to smaller radii. In addition, I find that this satellite-disk
interaction also builds a substantial surface density enhancement at
the disk edge. This may lead to enhanced star formation at the disk
edge.  This process which I call {\it gas shepherding} may be a
generic feature of gaseous disks around galaxies, if a sufficiently
massive and dense satellite is available.

The physics of satellite-disk interactions in galaxies or gas
shepherding is not just an interesting exercise in mathematical
physics, but may be important in the fueling of active galaxies and
central starbursts.  First, the action of dynamical friction on the
satellite-disk system will shrink radius of the disk, thereby
transporting gas to smaller radii.  This shepherding of gas will
continue until the gas is forced into the center or the shepherding
satellite is destroyed.  I will discuss this scenario for nuclear
fueling and how it compares to competing scenarios in this paper.

%Secondly, as I demonstrate below, the competition between
%disk tides and dynamical friction leads to a preferred length scale in
%the problem, which I call the ``Hill'' radius of the disk, which is an
%estimate for the separation between the disk and the satellite.  This
%may help explain the morphology of star cluster around nuclear rings.
%Finally, I argue that the physics of this may be especially relevant
%in the early universe as providing a means by which the first black
%hole may have been seeded.

I have structured this paper as follows.  In \S\ref{sec:basic
  picture}, I discuss the basic picture of gas shepherding and give a
few order of magnitude estimates.  I estimate the timescale for gas
shepherding and the expected surface density enhancement.  I present
the physics of gas shepherding and solve numerical models in
\S\ref{sec:shepherding}.  I also present an approximate analytic
solution to this problem.
%I discuss some applications this
%work in \S\ref{sec:application} including the effect of star
%formation on the shepherded disk and the destruction of the shepherd.
I then discuss the application of gas shepherding to the feeding of
active galactic nuclei and central starbursts in
\S\ref{sec:applications}.
% and the migration of star clusters in nuclear rings.  
Finally I present my conclusions in \S\ref{sec:conclusions}.

\section{Basic Picture}\label{sec:basic picture}

\begin{figure}
\plotone{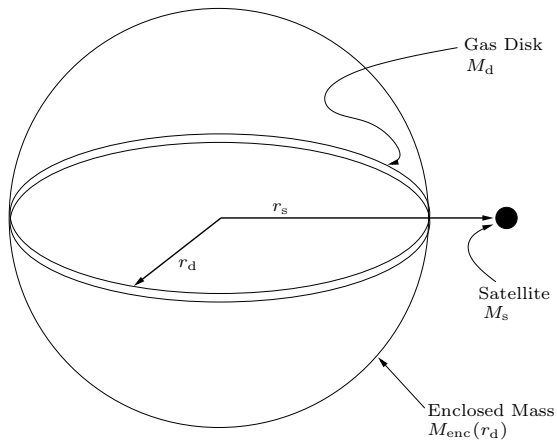}
\caption{Schematic of gas shepherding.  A satellite with mass, $M_{\rm
    s}$, and orbital radius, $r_{\rm s}$, infalls via dynamical
  friction and interacts with a gaseous disk with mass, $M_{\rm d} =
  \pi \Sigma r_{\rm d}^2$, that sits in a halo with total enclosed
  mass, $M_{\rm enc}$.}
\label{fig:cartoon1}
\end{figure}

The basic picture of gas shepherding is illustrated by Figure
\ref{fig:cartoon1}.  A gas disk with a radius of $r_{\rm d}$ and
surface density of $\Sigma$ orbits about a
spherical mass distribution with enclosed mass of stars or dark matter
of $M_{\rm enc}(r_{\rm d})$.  Due to dynamical friction on the stellar
background, an external satellite with mass $M_{\rm s}$ slowly spirals
in on a circular orbit with a radius of $r_{\rm s}$ in the same
orbital plane of the gas disk.  The radial velocity at which the
satellite initially spirals into the gas disk due to dynamical
friction is $v_{\rm df} \sim (M_{\rm s}/M_{\rm enc}) v_{\rm orb}$,
where $v_{\rm orb}$ is the orbital velocity of the satellite.  As the
satellite approaches the disk edge, i.e., $r_{\rm s}$ approaches
$r_{\rm d}$, the satellite excites spiral density waves at the
Lindblad resonances (Goldreich \& Tremaine 1980), which transfers
angular momentum from the disk to the satellite.  This angular
momentum is in turn transferred to the background stars or dark matter
via dynamical friction on the satellite.  Since angular momentum of
the satellite-disk system is continually bled, the radius of the
satellite-disk system will shrink at the shepherding velocity, $v_{\rm
  shep}$.

%The situation that
%I have presented is exceedingly simplified and I will discuss
%complications to this picture in \S\ref{sec:implications}.
%\clearpage
%\clearpage

I first give some simple order of magnitude estimates of this process.
The dynamical friction timescale for a satellite is (Chandrasekhar
1943; Binney \& Tremaine 1987)
\begin{equation}
t_{\rm df} \sim \frac {M_{\rm enc}}{M_{\rm s}}  t_{\rm dyn},
\end{equation}
where $t_{\rm dyn} = \Omega_{\rm s}^{-1}$ is the dynamical time,
$\Omega_{\rm s} = \sqrt{GM_{\rm enc}/r_{\rm s}^3}$ is the orbital
frequency of the satellite, and $G$ is Newton's constant.  The torque
due to dynamical friction on the satellite, $T_{\rm df}$, is
\begin{equation}\label{eq:torque estimate df}
  T_{\rm df} \sim M_{\rm s} r_{\rm s}^2 \Omega_{\rm s} t_{\rm df}^{-1}
  \sim \frac {G M_{\rm s}^2}{r_{\rm s}}
\end{equation}
The torque due to the excitation of spiral density waves in a disk is
(Goldreich \& Tremaine 1980; Lin \& Papaloizou 1986; Ward \& Hourigan
1989; Artymowicz 1993; Ward 1997)
\begin{equation}\label{eq:torque estimate disk}
  T_{\rm d} \sim \frac {G M_{\rm s}^2}{r_{\rm s}} \frac {M_{\rm d}}{M_{\rm enc}} 
  \left(\frac {r_{\rm s}} {r_{\rm s} - r_{\rm d}}\right)^3,   
\end{equation}
where $M_{\rm d} = \pi \Sigma r_{\rm d}^2$ is the mass of the disk, I
have presumed that radial separation between satellite and disk is
small compared to the radius, i.e., $r_{\rm s} - r_{\rm d} \ll r_{\rm
  s}$.

For now, I assume that there is no internal rearrangement of angular
momentum in the disk, i.e., an inviscid disk.  Dynamical friction
torques down the satellite ($T_{\rm df}$), while the disk, whose edge
sits at $r_{\rm d} < r_{\rm s}$, torques up the satellite ($T_{\rm
  d}$). When $T_{\rm df} \sim T_{\rm d}$, the satellite and the disk
are well coupled for $M_{\rm s} \le M_{\rm d}$, i.e., the mass of the
satellite is smaller or comparable to the mass of the disk.  Setting
$T_{\rm d} \sim T_{\rm d}$ gives
\begin{equation}
  \frac {r_{\rm s} - r_{\rm d}} {r_{\rm s}} \sim \left(\frac {M_{\rm d}}{M_{\rm enc}}\right)^{1/3}, 
\end{equation}
which I define as the ``Hill'' radius of the disk, $r_{\rm H,d} =
r_{\rm s} (M_{\rm d}/M_{\rm enc})^{1/3}$.\footnote{I use quotation
  marks because disks do not have Hill radii in the traditional sense.
  Rather I have adopted this terminology because of the familiar
  scaling, i.e., the 1/3 power law.} This gives the natural scale for the
radial separation between disk and satellite, i.e., $r_{\rm s} -
r_{\rm d} \sim r_{\rm H,d}$.

For spiral density waves that damp locally, the radial extent of the
disk over which the satellite exerts torques is also $\sim r_{\rm
  H,d}$.  Assuming no viscous spreading, the disk surface density will
be enhanced at the disk edge due to the piling up of material from the
initial radius of $r_{\rm d,0}$ to the current radius of $r_{\rm d}$.
This enhancement due to a satellite that sinks a distance $\sim r_{\rm d}$ is
\begin{equation}
  \frac {\Sigma}{\Sigma_0} \sim r_{\rm d}/r_{\rm H,d},
\end{equation} 
where $\Sigma$ is the surface density and $\Sigma_0$ is the initial
surface density. For $M_{\rm d}/M_{\rm enc} \sim 10^{-3} - 10^{-2}$,
the enhancement is a factor of a few to ten.

\section{Gas Shepherding}\label{sec:shepherding}

\subsection{Basic Equations}

To study this problem in greater detail, I begin by writing the
equations for a viscous disk that is evolving under the influence of
an external torque.  The equation of continuity is
\begin{equation}\label{eq:continuity}
\frac {\partial \Sigma} {\partial t} + \frac 1 r \frac {\partial
(r\Sigma v_r)}{\partial r} = 0,
\end{equation}
where $r$ is the radial coordinate, and $v_r$ is the radial component
of the velocity (Frank, King, amd Raine 2002).  The angular momentum
equation is
\begin{equation}\label{eq:angular momentum}
\Sigma\frac{\partial (r^2\Omega)}{\partial t} + v_r \Sigma \frac{\partial (r^2 \Omega)} {\partial r} = -\frac 1 {2\pi r}\left(\frac {\partial T_{\rm visc}}{\partial r} - \frac {\partial T_{\rm d}}{\partial r}\right),
\end{equation} 
where $\Omega = \sqrt{GM_{\rm enc}(r)/r^3}$ is the orbital frequency,
$M_{\rm enc}(r)$ is the mass enclosed inside $r$, and $T_{\rm visc} =
-2\pi r^3\nu\Sigma\partial \Omega/\partial r$ is the viscous torque,
$\nu$ is the viscosity (Frank, King, amd Raine 2002).  Using equation
(\ref{eq:angular momentum}), I find\footnote{The angular frequency,
  $\Omega$ is a function only of the coordinate, $r$, assuming that
  the mass enclosed, i.e., $M_{\rm enc}$ is not an explicit function
  of time.  Therefore, the first term completely disappears from
  equation (\ref{eq:angular momentum}).}
\begin{equation}\label{eq:radial velocity}
v_r = -\frac 1 {2\pi r\Sigma}\left(\frac{\partial (r^2\Omega)}{\partial r}\right)^{-1} \left[\frac {\partial}{\partial r}\left( T_{\rm visc} - T_{\rm d}\right)\right].   
\end{equation}
Plugging equation (\ref{eq:radial velocity}) into
(\ref{eq:continuity}), I find
\begin{equation}\label{eq:time dependent sigma}
\frac{\partial \Sigma}{\partial t} = \frac 1 {2\pi r} \frac {\partial}
{\partial r} \left(\frac{\partial (r^2\Omega)}{\partial r}\right)^{-1}
\left[\frac {\partial}{\partial r}\left( T_{\rm visc} - T_{\rm
d}\right)\right].
\end{equation}

For the torque density due to an orbiting satellite on a disk,
$\partial T_{\rm d}/\partial r$, I take a form suggested by Ward \&
Hourigan (1989) (see also Goldreich \& Tremaine 1980; Lin \&
Papaloizou 1979a, 1979b, 1986):
\begin{equation}
  \frac {\partial T_{\rm d}}
    {\partial r} = {\rm sgn}\left(r-r_{\rm s}\right)\beta \frac {G^2
    M_{\rm s}^2 \Sigma}{r(\Omega - \Omega_{\rm s})^2\left(r - r_{\rm s}\right)^2},
\end{equation}
where $\beta$ is a constant of order unity.  For $|r-r_{\rm s}| \ll r$,
I expand $\Omega-\Omega_{\rm s} \approx (\partial\Omega/\partial r) (r - r_{\rm s})$ to get
\begin{equation}
\frac {\partial T_{\rm d}}{\partial r} \approx {\rm
sgn}\left(r-r_{\rm s}\right)\beta \frac {G^2 M_{\rm s}^2 \Sigma r}{\Omega^2\left({r - r_{\rm s}}\right)^{4}},
\end{equation}
where I have taken a density profile for the enclosed mass of the form
$\rho\propto r^{-2}$, so that 
\begin{equation}\label{eq:mass profile}
M_{\rm enc}(r) = M_{\rm enc,0} \frac r {r_0},
\end{equation}
where $M_{\rm enc, 0}$ is the enclosed mass at $r_0$.
Note that for such a mass distribution, the orbital velocity, 
$v_{\rm orb} = \sqrt{G M_{\rm enc, 0}/r_0}$, is a constant.  
Using (\ref{eq:mass profile}), I find the following relations
\begin{equation}\label{eq:viscous torque}
\frac {\partial T_{\rm visc}}{\partial r} = 2\pi v_{\rm orb}\frac {\partial}{\partial r}\left(\Sigma \nu r\right),  
\end{equation}
\begin{equation}\label{eq:external torque}
\frac {\partial T_{\rm d}}{\partial r} \approx {\rm
sgn}\left(r-r_{\rm s}\right)\beta \frac {G M_{\rm s}^2 \Sigma r_0 r^3}{M_{\rm enc,0}\left({r - r_{\rm s}}\right)^{4}},
\end{equation}
Plugging equations (\ref{eq:viscous torque}) and (\ref{eq:external
torque}) into (\ref{eq:time dependent sigma}), I find
\begin{equation}\label{eq:time dep sigma}
  \frac{\partial \Sigma}{\partial t} = \frac{1}{r}\frac{\partial^2 \left(r\nu\Sigma\right)}{\partial r^2} + \frac {\beta}{2\pi}\frac {r_0} {v_{\rm orb} r} \frac {G M_{\rm s}^2}{M_{\rm enc,0}}\frac {\partial}{\partial r}\left[\frac {\Sigma r^3}{\left(r - r_{\rm s}\right)^{4}}\right],
\end{equation}
where I assume $r < r_{\rm s}$, which fixes the sign.  Note that the
term in equation (\ref{eq:radial velocity}) and (\ref{eq:time
  dependent sigma}), $\partial (r^2\Omega)/\partial r = v_{\rm orb}$,
is a constant using the prescribed mass distribution in equation
(\ref{eq:mass profile}).  I chose the viscosity law suggested by Lin
\& Papaloizou (1986):
\begin{equation}\label{eq:viscosity law}
  \nu = \nu_0\left(\frac{\Sigma}{\Sigma_0}\right)^2.
\end{equation}

The time-rate change of angular momentum of the satellite is
\begin{equation}
M_{\rm s}\frac {\partial  (r_{\rm s}^2 \Omega_{\rm s})}{\partial t} = -T_{\rm d} + T_{\rm df},  
\end{equation}
where $-T_{\rm d} = -\int dr \partial T_{\rm d}/\partial r$.  The
velocity of infall, i.e., the shepherding velocity, is $v_{\rm shep} =
\dot{r}_{\rm s}$.  Thus, I find
\begin{eqnarray}\label{eq:rdot}
  \frac {\partial r_{\rm s}} {\partial t} = \frac {GM_{\rm s}} {v_{\rm orb}}
\left[-\beta\int \frac {{\rm sgn}\left(r-r_{\rm s}\right)}{\left(r - r_{\rm s}\right)^4} \frac {\Sigma r_0 r^3}{M_{\rm enc,0}} dr 
    - \ln\Lambda\frac {1}{r_{\rm s}}\right].
\end{eqnarray}
Equations (\ref{eq:time dep sigma}) and (\ref{eq:rdot}) constitute a
complete set of equations which governs the behavior of the
satellite-disk system.  These equations are exactly the same as those
that govern the migration of protoplanets in protoplanetary nebula
(Lin \& Papaloizou 1979ab, 1986; Hourigan \& Ward 1984; Ward \&
Hourigan 1989; Ward 97; Rafikov 2002), except with the addition of
another term on the RHS of equation (\ref{eq:rdot}), which is due to
dynamical friction.

To simplify equations (\ref{eq:time dep sigma}) and (\ref{eq:rdot}), I
rescale the variables
\begin{eqnarray}
\sigma &=& \frac {\Sigma}{\Sigma_0}, \\
t' &=& \frac t {t_{\rm df}} =  \Omega_0\,t \ln\Lambda\frac {M_{\rm s}}{M_{\rm enc,0}},
\end{eqnarray}
where $\Sigma_0$ is the initial surface density of the disk, which I
assume to be constant and $\Omega_0$ is the orbital frequency at
$r=r_0$.  Equations (\ref{eq:time dep sigma}) and (\ref{eq:rdot})
become
\begin{eqnarray}
  \frac{\partial \sigma}{\partial t'} &=& \frac {\nu_0}{q\Omega_0 \ln\Lambda r} \frac {\partial^2(r\sigma^3)}{\partial r^2} +
{\beta'}\frac {q r_0^3}{r} \frac{\partial}{\partial r}\left(\frac {\sigma r^3}{(r-r_{\rm s})^4}\right), \\
\frac {\partial r_{\rm s}} {\partial t'} &=& 2\beta'\frac{\pi \Sigma_0 r_0^2}{M_{\rm enc,0}} r_0 \int \frac{\sigma r^3}{(r-r_{\rm s})^4} dr
-\frac {r_0^2}{r_{\rm s}}
\end{eqnarray}
where $q=M_{\rm s}/M_{\rm enc,0}$ and $\beta' =
\beta/(2\pi\ln\Lambda)$, and I have taken the viscosity to be $\nu =
\nu_0\sigma^2$ (see eq.[\ref{eq:viscosity law}]), where $\nu_0 =
\alpha (h_{\rm d, 0}/r_{\rm d,0})^2 r_{\rm d,0}v_{\rm
  orb}$,\footnote{Using the standard Shakura \& Sunyaev (1973) $\alpha$
  prescription for the viscosity, i.e. $\nu=\alpha c_s h$, where $c_s$
  is the sound speed and $h$ is the disk scale height, I write $c_s =
  v_{\rm orb} h/r$ and I take $h=h_{\rm d,0}$ at $r = r_{\rm d,0}$.}
where $r_{\rm d,0}$ is the initial radius, $h_{\rm d, 0}$ is the
initial vertical scale height of the disk, and $\alpha$ dimensionless
viscosity parameter (Frank, King, \& Raine 2002).  The initial scale
height of the disk can be written as $h_{\rm d, 0}/r_{\rm d,0} = Q_0
q_{\rm d}/\sqrt{2},$\footnote{To get this relation, take the Toomre Q
  parameter for a gaseous disk, $Q = c_s\kappa_0/\pi G\Sigma_0$
  (Binney \& Tremaine 1987).  The epicyclic frequency,
  $\kappa_0 = (4+2d\ln\Omega_0/d\ln r)^{1/2}\Omega_0 = \sqrt{2}\Omega_0$.
  Plugging $c_s = v_{\rm orb} h/r$, I find $h/r = (M_{\rm d}/M_{\rm
    enc})(Q/\sqrt{2})$.} where $Q_0$ is the initial Toomre Q (Toomre
1964) of the disk and $q_{\rm d} = M_{\rm d}/M_{\rm enc, 0}$ is the ratio
of the disk mass to the enclosed mass.  Without loss of generality, I
now take $r_0 = r_{\rm d,0} \approx r_{\rm s,0}$ to find
\begin{eqnarray}\label{eq:master1}
  \frac{\partial \sigma}{\partial t'} &=& \alpha' \frac {q_{\rm d}} {q'} \frac {r_{\rm d,0}^2} r \frac {\partial^2(r\sigma^3)}{\partial r^2} \nonumber\\ &+&
  {\beta'}\frac {q'q_{\rm d} r_{\rm d,0}^3}{r} \frac{\partial}{\partial r}\left(\frac {\sigma r^3}{(r-r_{\rm s})^4}\right), \\
  \frac {\partial r_{\rm s}} {\partial t'} &=& 2\beta'q_{\rm d} r_{\rm d,0} \int \frac{\sigma r^3}{(r-r_{\rm s})^4} dr
  -\frac {r_{\rm d,0}^2}{r_{\rm s}}\label{eq:master2}
\end{eqnarray}
where $q' = M_{\rm s}/M_{\rm d}$ and $\alpha'=\alpha Q_0^2/(2\ln\Lambda)$.
Equations (\ref{eq:master1}) and (\ref{eq:master2}) are the master
equations governing my model which I solve numerically in
\S\ref{sec:numerical} and solve approximately in \S\ref{sec:approx}.

\subsection{Numerical Results}\label{sec:numerical}

Equations (\ref{eq:master1}) and (\ref{eq:master2}) consists of four
dimensionless parameters: $q_{\rm d}$, $q'$, $\alpha'$ and $\beta'$.  For the
dimensionless parameters, $\beta'$, which represent the relative power
between disk tides and dynamical friction, I choose a fiducial value
for $\beta' = 1$, assuming fiducial values of
$\ln\Lambda \sim O(1)$ and $\beta \sim O(1)$.  For $\alpha'$, which
represents the relative power between viscosity and dynamical
friction, I take $\alpha' = 10^{-3}-10^{-1}$ to study the effects of
viscosity.  Finally, I set $q_{\rm d} = M_{\rm d}/M_{\rm enc} = 10^{-3}$. 
%so that $r_{\rm H,d}/r_{\rm d} = 0.1 \ll 1$.
  
I solve equations (\ref{eq:master1}) and (\ref{eq:master2}) using
standard explicit finite-difference methods (Press et al. 1992).  For
good spatial resolution, I select 750 grid points between $r/r_{\rm
d,0} = 0$ to $1.5$ with the satellite initially at $r_{\rm s} = 2
r_{\rm d,0}$.  The satellite begins to fall inward via dynamical
friction.  As the satellite approaches the disk, the disk exerts a
torque on the satellite, which slows its infall.  At the same time,
the surface density of the disk increases at the disk edge.

\begin{figure}
\plotone{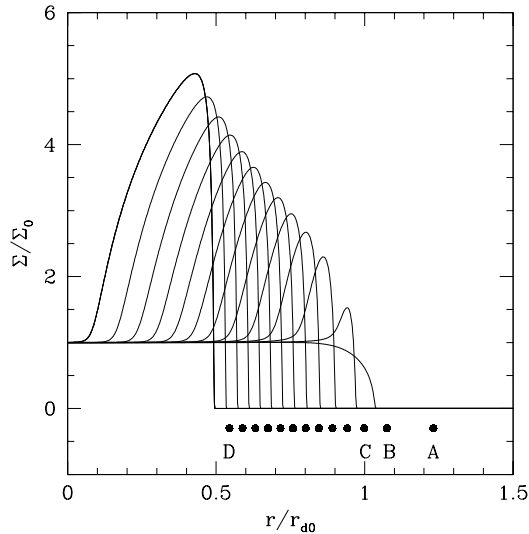}
\caption{Evolution of the surface density as a function of radius for
  a satellite disk system where $q' = 0.1$ for $\alpha'=0.1$ and
  $\beta'=1$. The snapshots are separated in $\Delta t = 0.25 t_{\rm
    df,0}$ increments. The satellite position is marked by the black
  dot below the surface density plot.  Each point and each curve
  correspond to one snapshot.  The labeled points, A-D, are explained
  in the text.}
\label{fig:md0.1}
\end{figure}

I show this evolution for $\alpha' = 0.1$, $\beta' = 1$, and $q' =
0.1$ in Figure \ref{fig:md0.1}.  I start the evolution of the
satellite at $r_{\rm s}/r_{\rm d,0} = 2$.  When the satellite sinks to
$r_{\rm s}/r_{\rm d,0} = 1.25$, I set $t=0$ and plot the evolution of
the satellite-disk system in intervals of $\Delta t = 0.25 t_{\rm
  df,0}$ for a total of $1.75 t_{\rm df,0}$ where $t_{\rm df,0}$ is
the dynamical friction timescale at $r_{\rm d,0}$.  At $t=0$ the disk
is undisturbed except for a small amount of spreading due to viscosity
at the disk edge.  The satellite position is at $r/r_{\rm d,0} \approx 1.25$
as noted by the black dot labeled A.  I evolve in intervals of $\Delta t$ and
note that the disk satellite system has begun to respond (point
labeled B). The disk begins to develop a surface density enhancement
and the satellite slows down a bit.  Evolving for another $\Delta
t$, i.e., the point labeled C, shows a substantial change.  The
surface density increases near the disk edge by a maximum factor
of $\approx 2$.  Note that the distance between point B and C is
substantially smaller than the distance between A and B, which results
from the backreaction of the disk tides.  By point D, the density
enhancement is $\approx 5$.  Note also that the distance between point
D and its neighbor is somewhat larger than the other separations.
This is due to the decrease in the dynamical friction timescale
compared to the dynamical timescale when the mass enclosed decreases
relative to the satellite mass.

%\clearpage
%\clearpage

\begin{figure}
\plotone{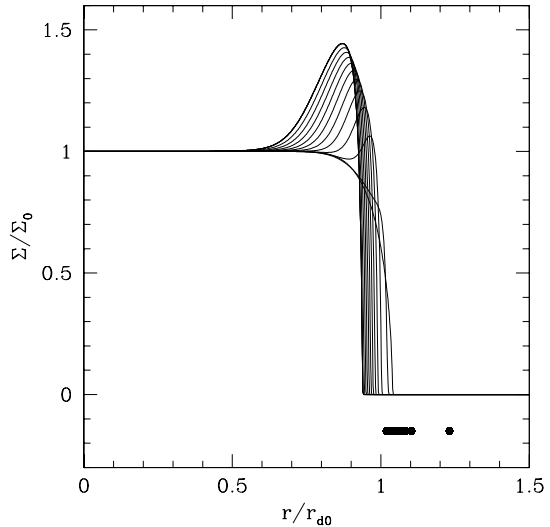}
\caption{Same as Figure \ref{fig:md0.1} but for $q' = 0.01$.  The
  larger mass of the disk stops the body from infalling toward the
  center.  In this case there is no surface density enhancement.}
\label{fig:md0.01}
\end{figure}

\begin{figure}
\plotone{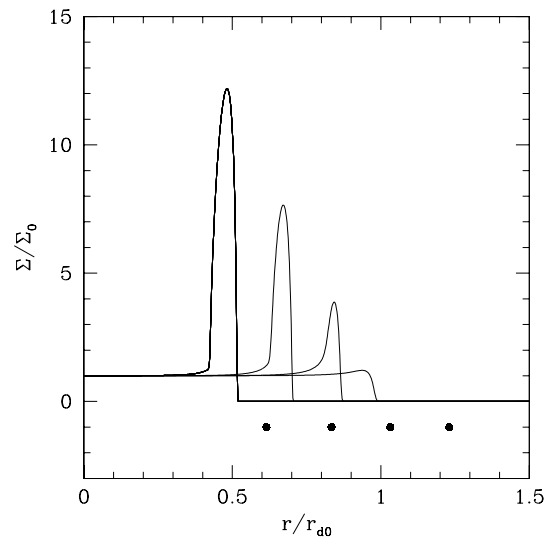}
\caption{Same as Figure \ref{fig:md0.1} but for $q' = 1$.  The surface
  density is enhanced over the case where $q' = 0.1$.  In addition the
  disk has a much reduced effect on the infall velocity of the
  satellite because they have equal masses.}
\label{fig:md1}
\end{figure}

In Figure \ref{fig:md0.01} and \ref{fig:md1}, I show the case for
$q'=0.01$ and $1$ respectively. For $q'=0.01$ (Fig. \ref{fig:md0.01}),
the initial infall of the satellite is stopped by the disk because of
the large mass ratio between the satellite and the disk.  In addition,
there is no surface density enhancement at the disk edge.  For $q'=1$
(Fig. \ref{fig:md1}), on the other hand, the initial infall of the
satellite is only slightly slowed because the mass of the disk and satellite 
are the same.  In this case, a huge surface density enhancement is
produced.

%\clearpage

%\clearpage

One remarkable aspect of Figures \ref{fig:md0.1} and \ref{fig:md1} is
that they appear to approach quasi-steady state solutions.  This is
especially clear in Figure \ref{fig:md0.1}. Motivated by the
appearance of these quasi-steady state solutions, I find an
approximate kinematic wave solution in \S\ref{sec:approx}, which
agrees well with these numerical results.

\begin{figure}
\plotone{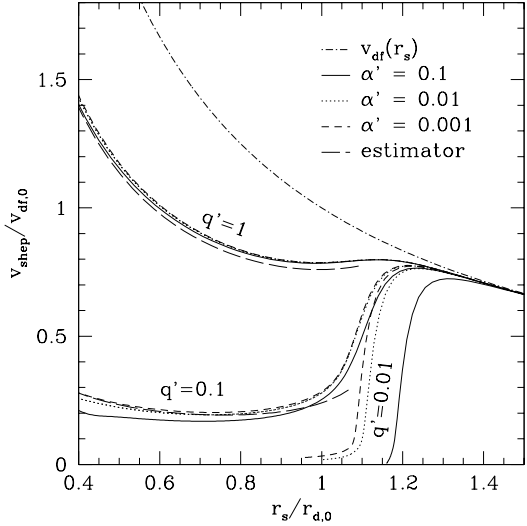}
\caption{Plot of the normalized shepherding velocity as a function of
  $r_{\rm s}$ for $q'=0.01$ (bottom set of curves), $0.1$ (middle
  set), and $1$ (top set) and $\alpha'=0.001$ (short-dashed lines),
  0.01 (dotted lines), and 0.1 (solid lines) normalized to the initial
  infall velocity, $v_{\rm df,0}$.  The local dynamical infall
  velocity, $v_{\rm df}$ (dot-dashed line), is also plotted for
  purposes of comparison.  Finally, the estimate for $v_{\rm shep}$
  (eq.[\ref{eq:vp estimate}]) is plotted for $r_{\rm s} < r_{\rm
    d,0}-r_{\rm H,d}$ to allow a kinematic wave solution to be
  established.}
\label{fig:vp}
\end{figure}

In Figure \ref{fig:vp}, I plot the normalized shepherding velocity of
the satellite disk system, i.e., normalized to the initial infall
velocity, $v_{\rm df,0} = v_{\rm df}(r_{\rm d,0})$ (lower plot) and
the local infall velocity, $v_{\rm df}$ (upper plot), as a function of
$r_{\rm s}$ for $q'=0.01$, 0.1, and 1.  I plot it for $\alpha'=0.001$,
0.01, and 0.1.  For comparison I plot the local dynamical friction
infall velocity, $v_{\rm df}(r_{\rm s})$.  For $r_{\rm s} > r_{\rm
  d,0}$, $v_{\rm shep} \approx v_{\rm df}$ because tidal coupling
between the disk and satellite has not yet been achieved.  Once the
coupling has been achieved, i.e. $r_{\rm s} < r_{\rm d,0}$, the
shepherding velocity depends mainly on the mass ratio between the
satellite and disk, $q'$, and is independent of the viscosity,
$\alpha'$.  This highlights the fact that dynamical friction dominates
the dynamics.  I also plot an estimate for migration velocity
(eq.[\ref{eq:vp estimate}]) for $r_{\rm s} < r_{\rm d,0}-r_{\rm H,d}$,
i.e., I allow the shepherding satellite to fall inside of one Hill
radius of the disk to allow a kinematic wave solution to be set up
(see \S\ref{sec:approx}).  This estimate (eq.[\ref{eq:vp estimate}])
gives remarkably good results, i.e., $\lesssim 10\%$ for the fidicial
cases of $q'=0.1$ and $1$.  The most remarkable aspect of Figure
\ref{fig:vp} is that while the infall velocity falls substantially, a
sizeable fraction of the initial infall velocity remains.  For
instance, the infall velocity does not fall below 20\% of the initial
infall velocity for the $q'=0.1$ case.

%\clearpage
%\clearpage

%The most remarkable aspect of Figure \ref{fig:vp} is that while the
%infall velocity falls substantially, a sizeable fraction of the
%initial infall velocity remains.  Intuitively, the infall should be
%slowed by a factor of the mass ratio, $q'$, which it approaches when
%compared to the {\it local} infall velocity, i.e., the upper plot of
%Figure \ref{fig:vp}.  While it does slow, two effects act to mitigate
%this effect.  First dynamical friction increases toward smaller radii
%because $v_{\rm df} \sim M_{\rm s}/M_{\rm enc}$ and $M_{\rm enc} \sim
%r$.  Secondly, the disk is at a smaller radius than the satellite,
%i.e., $r_{\rm s} - r_{\rm d} \sim r_{\rm H,d}$, so the specific
%angular momentum of material in the disk is smaller than that of the
%satellite (see \S\ref{sec:approx} for instance).
 
The shepherding satellite transports gas from large radii to smaller
radii by dumping the angular momentum of the gas into the background
stars or dark matter.  Moreover, this transport is achieved on a
dynamical friction timescale, which can be fast for sufficiently
massive satellites.  This transport will continue unless the gas is
consumed via star formation or the satellite is tidally disrupted by
the increasing density of the background (see
\S\ref{sec:shepherd destruction}).
 
\subsection{Approximate Solution}\label{sec:approx}

Motivated by the numerical results of \S\ref{sec:numerical} I now
solve equations (\ref{eq:master1}) and (\ref{eq:master2}) using a
kinematic wave approach inspired by the planetary literature (Hourigan
\& Ward 1984; Ward \& Hourigan 1989; Ward 1997; Rafikov 2002).  First
I approximate equations (\ref{eq:master1}) and (\ref{eq:master2}) to
recover a simpler set.  As the order of magnitude estimate in
\S\ref{sec:basic picture} reveals, the typical radial separation for a
well coupled satellite-disk system is the Hill radius of the disk,
$r_{\rm H,d}$.  Thus, I now change variables in equations
(\ref{eq:master1}) and (\ref{eq:master2}) from $r$ to $x$, where
\begin{equation}
x = \frac {r-r_{\rm s}} {r_{\rm H,d0}} = \frac {r-r_{\rm s}} {q_{\rm d}^{1/3} r_{\rm d,0}},
\end{equation}
where $r_{\rm H,d0} = q_{\rm d}^{1/3}r_{\rm d,0}$ is the {\em initial}
Hill radius of the disk.  Since $r_{\rm H,d0} \ll r$, I take the
approximation $r-r_{\rm s} \ll r$ to find
\begin{eqnarray}\label{eq:master approx1}
  \frac{\partial \sigma}{\partial t'} &=& \alpha'\frac {q_{\rm d}^{1/3}} {q'} \frac {\partial^2\sigma^3}{\partial x^2} +
\frac {q'} {q_{\rm d}^{2/3}}\beta'\left(\frac{x_{\rm s}}{x_{\rm d,0}}\right)^2 \frac{\partial}{\partial x}\left(\frac {\sigma}{x^4}\right),\\
\frac {\partial x_{\rm s}} {\partial t'} &=& x_{\rm d,0}\left[2\beta' \left(\frac {x_{\rm s}}{x_{\rm d,0}}\right)^3  \int \frac{\sigma}{x^4} dx
-\frac {x_{\rm d,0}}{x_{\rm s}}\right],\label{eq:master approx2}
\end{eqnarray}
where $x_{\rm s} = r_{\rm s}/r_{\rm H,d}$ and $x_{\rm d,0} = r_{\rm
  d,0}/r_{\rm H,d}$.\footnote{To aid in the derivation of equation
  (\ref{eq:master approx1}) and (\ref{eq:master approx2}), it is
  helpful to consider which terms will change rapidly for small $x$
  and which terms will change slowly. For instance, the term
  $(r-r_{\rm s})^{-4} \rightarrow x^{-4}$ changes rapidly with small
  $x$, whereas $r \rightarrow r = r_{\rm s} + r_{\rm H,d} x \approx
  r_{\rm s}$ changes very slowly.  Hence, I made the following series
  of substitutions $r\rightarrow r_{\rm s}$ and $r-r_{\rm s} \rightarrow
  r_{\rm H,d} x$.}

I now assume the kinematic wave {\it ansatz}: $\sigma = \sigma_{\rm w}(x -
v_{\rm shep} t')$, where $\sigma_{\rm w}$ denotes my kinematic wave
solution for $\sigma$.  Inserting this {\it ansatz} into equation
(\ref{eq:master approx1}), I find:
\begin{eqnarray}\label{eq:approx1}
-v_{\rm shep}\frac{\partial \sigma_{\rm w}}{\partial x} &=& \alpha'\frac {q_{\rm d}^{1/3}} {q'} \frac {\partial^2\sigma_{\rm w}^3}{\partial x^2}\nonumber\\ &+&
\frac {q'} {q_{\rm d}^{2/3}}\beta'\left(\frac{x_{\rm s}}{x_{\rm d,0}}\right)^2 \frac{\partial}{\partial x}\frac {\sigma_{\rm w}}{x^4},
\end{eqnarray}
Integrating once and ignoring the constant of integration, I find
\begin{equation}\label{eq:approx2}
-v_{\rm shep} = \frac {3\alpha'}{2} \frac {q_{\rm d}^{1/3}} {q'} \frac {\partial \sigma_{\rm w}^2}{\partial x} +
\frac {q'} {q_{\rm d}^{2/3}}\beta'\left(\frac{x_{\rm s}}{x_{\rm d,0}}\right)^2 \frac {1}{x^4},
\end{equation}
where I have cancelled one factor of $\sigma_{\rm w}$ from each term.  I now integrate once more to find 
\begin{equation}\label{eq:sigma approx1}
  \sigma_{\rm w}^2 = C + \frac {q'} {\alpha' q_{\rm d}^{1/3}}\left(-\frac 2 3 v_{\rm shep} x + \frac {2\beta'}{9} \left(\frac{x_{\rm s}}{x_{\rm d,0}}\right)^2\frac {q'} {q_{\rm d}^{2/3}} 
  \left(\frac {1} {x^{3}}\right)\right),
\end{equation}
where $C$ is the constant of integration.  The two terms in the
parenthesis are both negative because $x \propto r-r_{\rm s} < 0$ and
$v_{\rm shep} < 0$ (inward migration).  The first term in the
parenthesis cuts off the surface density as $x$ becomes more negative,
i.e., away from the disk edge, while the second term cuts off the
surface density as it approaches the disk edge due to tidal
interactions.  The peak, i.e., where $\sigma_{\rm w}^2$ is maximal, is
found from equation (\ref{eq:approx2}) by setting $\partial\sigma_{\rm
  w}^2/\partial x = 0$.  This occurs when\footnote{To clarify the
  signs in equation (\ref{eq:peak}), note that $x < 0$, i.e., the
  satellite is at larger radii than the gas and $v_{\rm shep} < 0$,
  i.e., the satellite-disk system moves inward.}
\begin{equation}\label{eq:peak}
x = x_{\rm peak} \equiv -\left(\frac {q'} {q_{\rm d}^{2/3}}\beta'\left(\frac{x_{\rm s}}{x_{\rm d,0}}\right)^2 \frac 1 {-v_{\rm shep}}\right)^{1/4}.
\end{equation}
Hence, $\sigma_{\rm w}(x_{\rm peak}) = \sigma_{\rm max}$, so given an estimate
for $\sigma_{\rm max}$, I can determine the constant of integration,
$C$, via
\begin{eqnarray}\label{eq:C determination}
  C &=& \sigma_{\rm max}^2 - \frac {q'} {\alpha' q_{\rm d}^{1/3}}\left(-\frac 2 3 v_{\rm shep} x_{\rm peak}\right.\nonumber\\ &+&\left. \frac {2\beta'}{9} \left(\frac{x_{\rm s}}{x_{\rm d,0}}\right)^2\frac {q'} {q_{\rm d}^{2/3}} 
  \left(\frac {1} {x_{\rm peak}^{3}}\right)\right),
\end{eqnarray}

In the appendix, I calculate $\sigma_{\rm max}$ by balancing the
torque from the satelite-disk interaction with the torque from
dynamical friction.  I find the following cubic equation
(eq.[\ref{eq:sigma max cubic app}]):
\begin{equation}\label{eq:sigma max cubic}
\frac {2\beta'}{3} \frac {\sigma_{\rm max}} {x_{\rm peak}^3}+2\left(\frac {x_{\rm d,0}}{x_{\rm s}}\right)^2 \frac {\alpha' q_{\rm d}} {q'^2}\sigma_{\rm max}^3 - \left(\frac {x_{\rm d,0}}{x_{\rm s}}\right)^4= 0,
\end{equation}
which I solve numerically.  This gives an estimate for $\sigma_{\rm
  max}$ and hence, $C$, given $v_{\rm shep}$.  Therefore, to complete
the calculation, I now give an estimate for $v_{\rm shep}$.  Before I
do so, I note that the RHS of equation (\ref{eq:sigma approx1}) can be
negative.  To join this kinematic wave solution, $\sigma_{\rm w}$,
onto the background solution, I set
\begin{equation}\label{eq:sigma approx}
  \sigma^2 = \left\{\begin{array}{ll}
      \sigma_{\rm w}^2 &{\rm if}\ \sigma_{\rm w}^2 > 0 \ {\rm and}\  x > x_{\rm peak} \\
      \sigma_{\rm w}^2 &{\rm if}\ \sigma_{\rm w}^2 > 1 \ {\rm and}\  x < x_{\rm peak}\\
      0 & {\rm if}\ \sigma_{\rm w}^2 < 0 \ {\rm and}\  x > x_{\rm peak} \\
      1 & {\rm if}\ \sigma_{\rm w}^2 < 1 \ {\rm and}\  x < x_{\rm peak}
      \end{array}\right..
\end{equation}

I now proceed to calculate $v_{\rm shep}$. I estimate the torque using
equation (\ref{eq:torque estimate df}).  The satellite raises tides on
the disk, which do not operate over the entire disk, i.e. the tides
are most acute at the edge of the disk due to the $(r-r_{\rm s})^{-4}$
scaling of the torque density. Rather they act on the accumulated mass
at the disk edge, $M_{\rm edge}$, which I estimate to be:
\begin{equation}\label{eq:mass edge}
M_{\rm edge} \approx M_{\rm d}\left[1 - \left(\frac {r_{\rm d,edge}}{r_{\rm d,0}}\right)^2\right],
\end{equation}
where $r_{\rm d,edge}$ is the edge of the disk.\footnote{In the
  nomenclature of this paper, $r_{\rm d,edge}$ should really be
  $r_{\rm d}$.  However, I will introduce this variable as I will fit
  for this and $r_{\rm H,eff}$.} To estimate $r_{\rm d,edge}$ for
equation (\ref{eq:mass edge}), I presume the following fit
\begin{equation}
r_{\rm d,edge} = r_{\rm s} - \psi r_{\rm H,d0},
\end{equation}
where $\psi$ is a constant of order unity and is determined below.
The total angular momentum of the satellite-disk system over which
these torques are effective is
\begin{equation}\label{eq:total L}
L = v_{\rm orb}\left(M_{\rm s} r_{\rm s} + M_{\rm edge} r_{\rm d}\right).  
\end{equation}
The difference between $r_{\rm d}$ and $r_{\rm s}$ is of order the
Hill radius of the disk, $r_{\rm H,d} = r_{\rm d}(M_{\rm d}/M_{\rm
  enc})^{1/3}$ (see \S\ref{sec:basic picture}).  As the satellite
falls inward and builds a large surface density enhancement, I expect
that the Hill radius of the disk will change, i.e., $\Sigma_0
\rightarrow \Sigma \approx M_{\rm edge}/4\pi r_{\rm d} r_{\rm H,d}$,
where I assume that the mass of the edge is spread over a radial
thickness of $r_{\rm H,d}$. Therefore, I define an effective Hill
radius of the disk by
\begin{eqnarray}
\frac {r_{\rm H,eff}}{r_{\rm d}} &=& \left(\frac {\pi \Sigma(r_{\rm d})r_{\rm d}^2}{M_{\rm enc}}\right)^{1/3}, \\
&=& \frac {r_{\rm H,d}}{r_0} \left(\frac {M_{\rm edge}}{2\pi M_{\rm d} r_{\rm H,d}}\right)^{1/3}.\label{eq:effective hill}
\end{eqnarray}
The disk edge is separated from the satellite by $\sim r_{\rm H,eff}$.
Thus, $r_{\rm d} \approx (1 - \eta r_{\rm H,eff}/r_{\rm d})r_{\rm s}$,
where $\eta$ is a constant of order unity.  Equating the time
derivative of $L$ (eq.[\ref{eq:total L}]) and (\ref{eq:torque estimate
  df}), I find
\begin{equation}
  \frac {GM_{\rm s}^2} {r_{\rm s}} \sim \dot{L} = v_{\rm orb}\left[M_{\rm s} + M_{\rm edge}\left(1 - \eta \frac {r_{\rm H,eff}}{r_{\rm d}}\right)\right] v_{\rm shep},
\end{equation}
where I ignore the time derivatives on $M_{\rm edge}$.  This gives
% (with the use of eq.[\ref{eq:effective hill}])
\begin{eqnarray}\label{eq:vp estimate}%\label{eq:vp estimate1}
  v_{\rm shep} &\sim& v_{\rm orb}\frac {M_{\rm s}}{M_{\rm enc}(r_{\rm s})} \frac {M_{\rm s}}{M_{\rm s} + M_{\rm edge}\left(1 - \eta{r_{\rm H,eff}}/{r_{\rm d}}\right)}.
\end{eqnarray}
Despite it crudeness, this velocity estimator gives reasonably good
agreement ($\lesssim 10\%$) when I select $\eta = 1$ and $\psi = 2$
via trial and error.  I plot the comparison in Figure \ref{fig:vp}.

\begin{figure}
\plotone{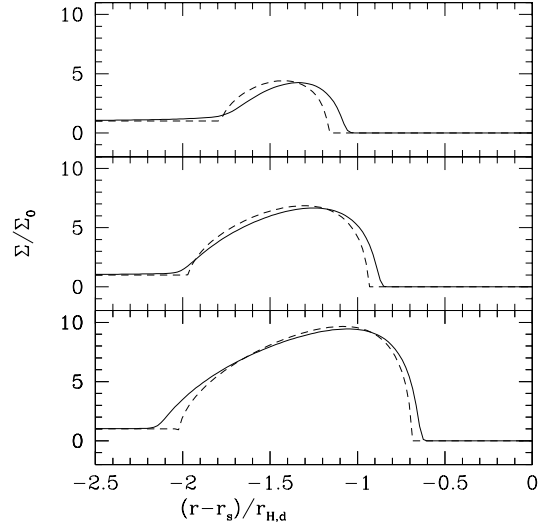}
\caption{The surface density enhancement $\Sigma/\Sigma_0$ as a
  function of $(r-r_{\rm s})/r_{\rm H,d}$ for the approximate solution
  (dashed lines) and the numerical solution (solid lines) for a
  satellite disk system where $q' = 0.1$ for $\alpha'=0.1$ and
  $\beta'=1$. The snapshots are separated in $\Delta t = 0.5 t_{\rm
    df,0}$ increments and the corresponding position of the
  shepherding satellite (from top to bottom) is $r_{\rm s}/r_{\rm d,0}
  = 0.95$, $0.75$, and $0.65$.}
\label{fig:analytic_md0.1}
\end{figure}

Equations (\ref{eq:sigma approx1}), (\ref{eq:C determination}),
(\ref{eq:sigma max cubic}), (\ref{eq:sigma approx}), and (\ref{eq:vp
estimate}) constitute a complete analytic solution once the evolution
of the disk can be modeled as kinematic wave, i.e., after the
satellite sinks down a few $r_{\rm H,d}$.  This approximate solution
agrees well with the exact numerical solution as shown in Figure
\ref{fig:analytic_md0.1} for $q' = 0.1$.  The agreement for $q'=1$ and
$q'=0.01$ is not as good.  For $q'=1$, the assumption that $T_{\rm d}
\approx T_{\rm df}$ breaks down and the agreement can be improved by
estimating $T_{\rm d} \approx T_{\rm df}/(1+q')$.  For $q'=0.01$,
kinematic wave solutions are not established because the infall of the
satellite is stopped.
%This solution can be
%improved with a better estimate for $v_{\rm shep}$ (eq.[\ref{eq:vp
%estimate}]).  Using the numerically calculated $v_{\rm shep}$ from the
%full numerical solution, the agreement  the approximate
%solution and exact numerical solution is excellent.

%\clearpage
%\clearpage

Finally, from the estimate for $v_{\rm shep}$ given in equation
(\ref{eq:vp estimate}) and Figure \ref{fig:vp}, I note $v_{\rm shep}$
is a factor of a few larger than $(M_{\rm s}/M_{\rm d}) v_{\rm df}$,
where $v_{\rm df}$ is taken at $r_{\rm d,0}$.  Hence I estimate the
timescale for shepherding to be $t_{\rm shep} \sim f (M_{\rm d}/M_{\rm
s}) t_{\rm df}$ or
\begin{equation}\label{eq:shepherding timescale}
t_{\rm shep} \sim f \frac {M_{\rm d}}{M_{\rm s}} \frac {M_{\rm enc}}{M_{\rm s}} \frac {t_{\rm dyn}}{\ln\Lambda},
\end{equation}
where $f < 1$ and $t_{\rm df} \sim ({M_{\rm enc}}/{M_{\rm s}}){t_{\rm
    dyn}}/{\ln\Lambda}$.

%\begin{figure}
%\plotone{analytic_md0.1.eps}
%\caption{The surface density enhancement $\Sigma/\Sigma_0$ as a
%  function of $(r-r_{\rm s})/r_{\rm H,d}$ for the approximate solution
%  (dashed lines) and the numerical solution (solid lines) for a
%  satellite disk system where $q' = 0.1$ for $\alpha'=0.1$ and
%  $\beta'=1$. The snapshots are separated in $\Delta t = 0.5 t_{\rm
%    df,0}$ increments and the corresponding position of the
%  shepherding satellite is $r_{\rm s}/r_{\rm d,0} = 0.95$, $0.79$,
%  $0.63$, and $0.46$.}
%\label{fig:analytic_md1}
%\end{figure}

%\section{Applications}

%The generic properties of satellite-disk interaction in galaxies,
%i.e., gas shepherding, are a separation between the satellite and disk
%of order one ``Hill'' radius of the disk, a radial shrinking of the
%satellite-disk system provided that the satellite is sufficiently
%massive (dynamical friction operates on a sufficiently rapid
%timescale), and a buildup of mass at the disk edge.  Each of these
%properties may be suitable for application to a variety of phenomenon
%in galaxies.  Thus far, I have identified two such applications of
%this physics: fueling of AGNs and central starbursts and migration of
%star clusters in nuclear ring systems.  I now discuss these two
%applications below. 

\section{The Fueling of Central Starbursts and Active Galactic Nuclei}\label{sec:applications}

%\subsection{Introduction to the problem}

As the satellite-disk system should continually shrink due to
dynamical friction, it raises the interesting question of whether or
not such a mechanism may be deliver gas into the nuclear regions of
galaxies.  The fueling of central starbursts and active galactic
nuclei remains an open question.  It is generally believed that the
fueling of central starbursts and active galactic nuclei (AGNs)
requires the delivery of gas from large radii to small radii (for a
review see Shlosman, Begelman, \& Frank 1990; hereafter SBF90).  Gas
transport is mediated at small radii ($\sim 1$ pc) by a thin
accretion disk.  At much larger radii ($\gtrsim 0.1 - 1$ kpc),
galactic bars can efficiently transport gas.

How is gas transported in the gap between $0.1-1$ kpc to $\sim 1$ pc?
Galactic bars, while efficient at large scales, stop at the inner
Lindblad resonance (ILR), which is at $\sim 0.1 - 1$ kpc.\footnote{As
  pointed out by the referee, ILRs may or may not exist, depending on
  the nuclear mass concentration.  However, these bars still become
  inefficient at small scales, i.e., typically 10\% of the size of the
  bar (Shlosman, Frank, \& Begelman 1989).} Thin accretion disks,
which are sufficient at small scales, would become gravitationally
unstable and would fragment to form stars outside of $\sim 1$ pc
(Paczynski 1978; Kolykhalov \& Sunyaev 1980; Shlosman \& Begelman
1987; Goodman 2003; Thompson, Quataert \& Murray 2005).

There are several possible modes by which this transport may be
accomplished.  First the gas is transported by large-scale {\it
  gaseous} gravitational instabilities, i.e., the ``bars within bars''
model (Shlosman, Frank, \& Begelman 1989).  Secondly, the gas may form
a thick turbulent marginally stable accretion disk powered by star
formation (Paczynski 1978; Kolykhalov \& Sunyaev 1980; Shlosman \&
Begelman 1987; Goodman 2003) via radiation pressure (Goodman \& Tan
2004; Thompson et al.  2005) or supernova (Wada \& Norman 2002).
Third, if the gas exist in the form of molecular clouds, collisions
between them may lead to episodic feeding of the nuclear regions,
i.e., the chaotic accretion scenario (Krolik \& Begelman 1988; SBF90;
Hopkins \& Hernquist 2006; King \& Pringle 2007; Nayakshin and King
2007).  Finally, the fueling of these central regions by be one of
brute force, where major or minor mergers drive gas from the disk of
the galaxy to the nuclear region (Hernquist 1989; Bekki \& Noguchi
1994; Mihos \& Hernquist 1994; Hernquist \& Mihos 1995)

In the ``bars within bars'' scenario, Shlosman, Frank, \& Begelman
(1989) argued that if a gravitationally unstable gas disk is
sufficiently massive such that it fulfills the Ostriker-Peebles
criterion (Ostriker \& Peebles 1973), it forms a secondary gaseous bar
which funnels gas inward on a dynamical time.  Numerical simulations
confirmed the viability of ``bars within bars'' (see for instance
Friedli \& Martinet 1993; Levine et al 2007).  It demands a
substantial mass in gas, $M_{\rm gas}\sim M_{\rm enc}$, where $M_{\rm
  enc}$ is the enclosed mass. While such large gas concentrations are
observed (see for instance Jogee et al.  2005), delivered into the
nuclear region by large scale bars, it is unclear if this mechanism
drives the bulk of AGN fueling and quasar activity.  Namely, the
observed correlation between large-scale bars, which transports gas
into the central region of galaxies, and Seyfert activity is
statistically marginal (see the review by Ho 2008 and references
therein; see the review by Combes 2003 and references therein; but
also see Laine et al 2002).  In addition, there is no evidence that
Seyferts have a higher fraction of nuclear bars (see Knapen 2004 and
references therein).\footnote{As pointed out by the first referee,
  however, the nuclear bars may be very short lived and therefore, may
  be very difficult to observed.}

On the other hand, if the disk is gravitationally unstable but lacks
sufficient mass to fulfill the demands of ``bars within bars'', the
gas will fragment and form stars. The process of star formation may
feed back on the disk to provide vertical support, maintaining
marginal stability, i.e., $Q\sim 1$, where $Q$ is the Toomre Q (see
for instance Shlosman \& Begelman 1989; Goodman 2003; Goodman \& Tan
2004; Thompson et al. 2005).  These thick marginally stable disks may
have a sufficiently large scaleheight such that accretion via viscous
processes is possible before all the gas is consumed.

Third, perhaps the gas exist primary in the form of molecular clouds.
Collisions between these molecular clouds provides something akin to a
viscosity (Krolik \& Begelman 1988; SBF90).  These
clouds would then feed the nuclear region in a chaotic fashion
(Hopkins \& Hernquist 2006; King \& Pringle 2007; Nayakshin and King
2007).  Typically, the viscosity due to cloud-cloud collision appears
to be insufficient for fueling the central regions from hundreds of
parsecs. However, it may be necessary around the sphere of influence
of the central black hole where large scale instabilities may not
operate (SBF90; see \S\ref{sec:final 10 pc})

Fourth, the fueling of central nuclear activity need not be due to gas
in the ISM, but gas liberated from disrupted stars.  Hills (1975)
proposed that the tidal disruption of stars (Rees 1988) may be a means
by which active galaxies may be fueled. These rates of tidal
disruption are enhanced in mergers of galaxies (Roos 1981), which
links the three phenomenon of merging galaxies, active galactic
nuclei, and tidal disruption of stars.  Alternatively, a pre-existing
gas disk may capture stars plunging through it, subsequently
disrupting these stars, which replenishes the disk.  This gas may then
viscously accrete, fueling the central engine (Miralda-Escud{\'e} \&
Kollmeier 2005).

Finally, mergers between galaxies may drive gas toward the nuclear
regions of galaxies where they fuel nuclear activity (Hernquist 1989;
Bekki \& Noguchi 1994; Bekki 1995).  This possibility may fuel
ultraluminous infrared galaxies and quasar activity as modeled, for
instance, by Springel et al. (2005) and Hopkins et al. (2005) and has
significant observational support (see for instance Sanders et al
1988). However, active galactic nuclei (AGN) activity appears to be
due to more quiescent phenomenon (Li et al 2006; Hopkins \& Hernquist
2006).  For instance, Li et al. (2006) explored the clustering
properties of AGN in Data Release 4 of the Sloan Digital Sky Survey
(SDSS) and concluded that interactions are unlikely to be the main
culprit in the fueling of AGNs.  Minor mergers are another possibility
(see for instance Mihos \& Hernquist 1994; Hernquist \& Mihos 1995;
Tanguchi \& Wada 1996; Taniguchi 1997, 1999).

\subsection{Fueling via Gas Shepherding}

Gas shepherding is another scenario by which gas is tranported from
hundreds of parsecs.  The gas shepherding scenario, which I
present is as follows: large scale bars drive gas toward the central
regions of a galaxy where they collect in a nuclear ring, i.e. the gas
transitions from $X_1$ to roughly circular $X_2$ orbits (Binney \&
Tremaine 1987) at a radii of a few hundred parsecs to 1-2 kiloparsecs
(Buta \& Combes 1996; Knapen 2004).  This gaseous ring becomes
gravitationally unstable and forms giant molecular clouds which in
turn collapse to form star clusters in coplanar orbits (see for
instance, Maoz et al.  2001; Jogee et al.  2002).  If the star cluster
formed is sufficiently massive, it may shepherd in the remnant gas in
the gaseous ring from a few hundred parsecs into the center.

I first compare the timescale of shepherding and viscous processes.
For an $\alpha$ viscosity, the viscous timescale is $t_{\rm visc} \sim
\alpha^{-1} \left(r/h\right)^2 t_{\rm dyn}$ (Frank, King, \& Raine
2002).  For a marginally stable disk, i.e., $Q\sim 1$, $h/r \sim
M_{\rm d}/M_{\rm enc}$.  Therefore the ratio between the shepherding
timescale, $t_{\rm shep}$, and the viscous timescale, $t_{\rm visc}$,
is
\begin{eqnarray}
\frac {t_{\rm shep}}{t_{\rm visc}} &\sim& \frac {f\alpha}{\ln\Lambda} \frac {M_{\rm d}} {M_{\rm enc}} \left(\frac {M_{\rm d}}{M_{\rm s}}\right)^2,\nonumber\\
&\sim&10^{-3} \left(\frac {\alpha}{0.1}\right) \left(\frac {f/\ln\Lambda}{0.1}\right) \left(\frac{M_{\rm d}/M_{\rm s}}{3}\right)^2\nonumber\\ &&\left(\frac {M_{\rm d}/M_{\rm enc}}{0.01}\right).\label{eq:visc_vs_shep}
\end{eqnarray}
These timescales become comparable to each other for thicker disks,
i.e. $h/r \sim M_{\rm d}/M_{\rm enc} \sim 1$, or lower mass
satellites, i.e., $M_{\rm d}/M_{\rm s} \sim 100$ or a combination of
the two. Thus when shepherding operates, it dominates over viscous
processes for low mass disk and large satellites.  Therefore, for this
scenario to work, the star cluster that is formed must be sufficiently
massive compared to the mass of the gas disk.  I would expect it to be
at least 10\% or greater in order for shepherding to be efficient
based on equation (\ref{eq:visc_vs_shep}) and on my calculations in
\S\ref{sec:shepherding}.  

If gas shepherding operates, what ultimately happends to this
shepherded gas?  First, if this gas is in a marginally stable disk, it
may fragment into stars at the disk edge (see for instance Thompson et
al. 2005; Levin 2007; Chang et al.  2007; Nayakshin, Cuadra, \&
Springel 2007; Van der Ven \& Chang in preparation).  Secondly, this
gas may be shepherded into a radius where the shepherding satellite is
tidally disrupted.  I now discuss these two possibilities below.

\subsection{Disk Fragmentation and Star Formation}\label{sec:star formation}

Marginally stable disks with sufficiently rapid cooling ($t_{\rm cool}
< 3\,t_{\rm dyn}$) will tend to fragment (Gammie 2001).  In the
galactic context this fragmentation will lead to star formation, which
may feed back on the disk, maintaining marginal stability.  This
feedback may be due to momentum driving (Murray, Quataert, \& Thompson
2005; Thompson et al. 2005), cosmic ray feedback (Socrates, Davis, \&
Ramirez-Ruiz 2008), or supernovae driven turbulence (Wada \& Norman
2002).  A surface density enhancement of a few to ten would enhance
local star formation.  I take the standard Kennicutt law (Kennicutt
1998) for the star formation rate to be
\begin{equation}\label{eq:starformationrate}
  \dot{\Sigma}_* = \Sigma \Omega \eta,
\end{equation}
where $\dot{\Sigma}_*$ is the star formation rate per unit area,
$\Sigma$ is the gas surface density, $\eta \sim 0.01$ is the
empirically determined efficiency.  Equation
(\ref{eq:starformationrate}) suggests that the star formation rate
would also be enhanced by the same factor as that of the surface
density.

Equation (\ref{eq:starformationrate}) implies that the timescale for
the gas to fragment into stars is
\begin{equation}\label{eq:sf timescale}
t_* \sim \frac {\Sigma}{\dot{\Sigma}_*} = \eta^{-1} t_{\rm dyn}.  
\end{equation}
The ratio between the timescale for shepherding and the timescale for
the gas to fragment into stars is 
\begin{eqnarray}\label{eq:shep vs sfr}
  \frac {t_{\rm shep}}{t_*} &\sim& \eta \frac f {\ln\Lambda} \left(\frac {M_{\rm d}}{M_{\rm s}}\right)^2\frac {M_{\rm enc}}{M_{\rm d}}\\
&\sim& 1 \left(\frac {\eta}{0.01}\right) \left(\frac {f/\ln\Lambda}{0.1}\right)\left(\frac{M_{\rm d}/M_{\rm s}}{3}\right)^2\nonumber\\ && \left(\frac {M_{\rm enc}/M_{\rm d}}{100}\right).
\end{eqnarray}
Thus, sufficiently large satellites can shepherd a considerable
fraction of the gas into the central regions before the gas completely
fragments into stars.  For typical numbers ($M_{\rm enc} \sim
10^9\,{\rm M}_{\odot}$), this implies a satellite mass $\sim
10^6\,{\rm M}_{\odot}$.

For less massive satellites, the disk would completely fragment into
stars.  Thus, the evolution would likely be different.  As the
satellite continues to fall inward, it captures these newly-formed
stars into mean-motion and secular resonances, which increases the
star's eccentricity and inclination.  Indeed, Yu, Lu, \& Lin (2007)
have suggested that this exact process may be responsible for the
distribution of the young stars in the Galactic Center (GC).  In this
case, an intermediate-mass black hole (IMBH) or dark cluster falls
into a disk of stars in the same orbital plane.  The stellar disk is
heated up as the stars are captured into the previously-mentioned
resonance.  Some of these stars are forced to migrate with the
infalling IMBH, while others are forced into close encounters.  Some
of these stars that are subjected to these close encounters may be
ejected out of the GC at large velocities, producing hypervelocity
stars (HVSs) or hypervelocity binaries (HVBs) (Lu, Yu, \& Lin 2007).
Hence stars, whose semi-major axis is larger than the current radius
of the infalling IMBH would be excited into a torus-like structure,
while stars interior to the radius of the infalling IMBH would remain
in a cold disk.  Such a dynamical distribution may result from the
effects of shepherding, but at much larger scales, i.e., $\sim 100$ pc
rather than $\sim 0.4$ pc as in the case in the GC.

%The presence of gas would complicate matters.  The gaseous component
%would damp the growth of eccentricity and inclination of the newly
%formed stars as it would in the protoplanetary context (REF).
%Therefore it is unclear how a fragmenting stellar disk or ring would
%evolve in the presence of a shepherding body.

\subsection{Destruction of the Shepherd}\label{sec:shepherd destruction}

If this gas does not fragment into stars (which is the case for
sufficiently massive satellites), gas shepherding may transport a
substantial amount of gas from $\sim 0.1$ kpc to smaller radii.  As
long as the satellite remains coherent, there is no limit to how far
gas may be shepherded.  This requirement that the satellite be
sufficiently massive ($\sim 10^6\,{\rm M}_{\odot}$)
limits the number of possible systems that can serve as a shepherd.

A super star cluster (SSC) is one such candidate.  SSCs range up to
mass of $\sim 10^6 - 10^7\,{\rm M}_{\odot}$ with a radius of parsecs
(see for instance McCrady \& Graham 2007; Hagele et al.  2007) so they
are sufficiently massive to transport gas before it completely
fragments into stars.  However, as they sink deeper into the bulge,
tidal forces may disrupt them, stopping the shepherding process.  The
SSC is tidally disrupted when the enclosed background density
$\rho_{\rm enc} \propto r^{-2}$ is equal to their mean stellar
density, $\rho_{\rm SSC} \sim M_{\rm SSC}/a_{\rm SSC}^3$, where
$a_{\rm SSC}$ is the radial size of the SSC.  Plugging in a few
example numbers, I find that $\rho_{\rm SSC} \sim 10^6 (M_{\rm
  SSC}/10^6\,{\rm M}_{\odot}) (a_{\rm SSC}/1\,{\rm pc})^{-3}\,{\rm
  M_{\odot}\,pc}^{-3}$.  For typical numbers, i.e., $M_{\rm enc} \sim
10^9\,{\rm M}_{\odot}$ at $100$ pc, $\rho_{\rm enc} \sim
10^{7}\,(r/1\,{\rm pc})^{-2} \,{\rm M_{\odot}\,pc}^{-3}$ so the radius
where the satellite is disrupted is
\begin{equation}\label{eq:tidal disruption}
  r_{\rm tid} \sim 12 \left(\frac {M_{\rm SSC}} {5 \times 10^5\,{\rm M}_{\odot}}\right)^{-1/2}
  \left(\frac {a_{\rm SSC}}{2\,{\rm pc}}\right)^{3/2}\,{\rm pc},
\end{equation}
which would also be where the gas disk is shepherded down to.  Note
that I have ignored the effect of the SMBH, which begins to become
important at such radii.  The densities and density structures of
these SSCs are not very well constrained.  McCrady and Graham (2007)
presents a very detailed study of the size and mass distrubution of
super star clusters in the nearby starbursting galaxy M82.  The mean
half power radius of these SSCs is 1.8 pc (McCrady and Graham 2007,
their Figure 7), and there are quite a few clusters around $5\times
10^5\,{\rm M_{\odot}}$. By equation (\ref{eq:tidal disruption}), this
gives a tidal disruption radius of about 10 pc.  Lets consider two
examples in more detail.  One example is SSC L in M82 which has a
virial mass of $\approx 4\times 10^6\,{\rm M_{\odot}}$, a projected
half-light radius of $\approx 1.5\,{\rm pc}$, and a tidal disruption
radius of $r_{\rm tid} \approx 3\,{\rm pc}$. Another example is SSC
11, which has a virial mass of $\approx 4\times 10^5\,{\rm
  M_{\odot}}$, a projected half light radius of $\approx 1.1$ pc, and
a tidal disruption radius of $\approx 6$ pc.  In another study, Hagele
et al (2007) observed five star forming regions of the circumnuclear
ring of NGC 3351 and found star cluster masses between $1.8-8.9\,{\rm
  M}_{\odot}$ and radii of $1.7-4.9$ pc. This would bracket tidal
disruption radii between a few parsecs to 25 parsecs.  Hubble space
telescope studies by Maoz et al (1996) suggest that the star clusters
formed in circumnuclear rings have sizes $<5$ pc.  Hence, SSCs appears
have sufficient mass and density to shepherd gas into ten to tens of
parsecs.

The destruction of a SSC may not be the result of tidal processes, but
instead to internal processes, i.e., stellar mass loss.  Thus, 
SSCs may have a short lifetime in analogy with stellar clusters in our
galaxy, i.e., open clusters and OB associations.  For typical galactic
clusters, they are disrupted because gas ejection from the formation
of HII regions and supernova (Lada and Lada 2003).  However, SSCs are
thought to be the progenitors of globular clusters, whose lifetime is
$\sim 10^{10}$ years, and thus their lifetime may be substantially
longer.  The vulnerability of SSCs to these internal processes
strongly depend on their stellar initial mass functions, i.e., they
must have a sufficient number of low mass stars (Meurer et al 1995).
Top-heavy IMFs are vulnerable to the disruptive processes of stellar
mass loss and other processes (Chernoff \& Weinberg 1990).  Young,
dense stellar clusters, the closest local analogues to SSCs, show a
significant population of low mass stars.  For instance, the $2\times
10^4\,{\rm M}_{\odot}$ (Walborn et al 2002) stellar cluster, R136 in
the Large Magellanic Cloud has a large population of low mass stars
down to $0.1\,{\rm M}_{\odot}$ (Brandl et al 1996).  In addition,
McCrady et al. (2003) found SSC 9 was consistent with a normal IMF,
though SSC 11 appeared to have a top heavy IMF.

%This heating is extremely small.  If I plug in typical numbers
%into equation (\ref{eq:heating}), I find
%\begin{equation}
%\dot{E} \sim 10^{34} \left(\frac {M_{\rm s}}{10^6\,M_{\odot}}\right)^2 
%\left(\frac {M_{\rm enc}} {10^9\,M_{\odot}}\right)^{1/2}\left(\frac {r_{\rm s}} {100\,{\rm pc}}\right)^{-5/2} \,{\rm ergs\ s^{-1}}
%\end{equation}

%Another possibility, which was point out to me by Jeremy Goodman, are
%the remnants from minor mergers of dwarf galaxies.  The dense cores which  would

%\subsection{More Fundamental Extensions}\label{sec:inclination}

Aside from SSCs, the nucleated remains of an accreted satellite galaxy
may also be the shepherds that herds that gas into the central regions
of the galaxy for sufficiently small mass ratios.  For mass ratio
between the mass of the satellite galaxy's remnant and the {\it
  dynamical mass} is of order 10\% or greater, long range tidal
interactions may disrupt a central disk before the shepherding
scenario outlined aboved can be set up and would be akin to the minor
merger scenario studied previously (Hernquist 1989; Bekki \& Noguchi
1994; Mihos \& Hernquist 1994; Hernquist \& Mihos 1995; Tanguchi \&
Wada 1996; Taniguchi 1997, 1999).  Using the nucleated remains of
satellite galaxies as shepherds is interesting, but a detailed study
of this scenario is left to future work.

\subsection{The Final Ten (or Tens of) Parsecs}\label{sec:final 10 pc}

As I argue in \S\ref{sec:shepherd destruction}, SSCs can shepherd gas
to within 10 to 10s of parsecs of the center.  However, due to tidal
disruption of the shepherds, gas shepherding is less effective at
these small radii as a mechanism to drive gas further inward.  A
similar problem exists for the ``bars within bars'' scenario.  Once
gaseous bars funnel the gas within the gravitational sphere of
influence of a black hole, it becomes increasingly difficult to have a
sufficient amount of mass to form nested bars (SBF90 their Figure 2).
As a result, another means of gas transport between ten to tens of
parsecs must operate for the gas shepherding and ``bars within
bars'' scenarios, i.e., the ``10 parsec'' problem.  One possibility,
outlined by SBF90, are cloud-cloud collisions.  At the deposition
radius, the gas becomes marginally self-gravitating and fragments into
molecular clouds whose collisions with one another serves as a source
of viscosity which drives material into the central regions.
Typically, the viscosity scales like:
\begin{equation}
\alpha_{\rm cl-cl} = \left\{\begin{array}{ll}
      C &{\rm for}\ C \ll 1\\
      C^{-1} &{\rm for}\ C \gg 1
      \end{array}\right.,
\end{equation}
where $C$ is the filling fraction of the clouds (see SBF90).  It is
unclear if this is more effective than magnetic (see the review by
Balbus 2003) or gravitationally driven viscosities (Gammie 2001).
%However, the gravitational scattering between clouds, collisions, or
%gravitational instabilities may heat the disk of clouds up to $h/r\sim
%0.3$ or (in the nomenclature of SBF90) $\Delta v/v \sim 0.3$.

The effects of cloud-cloud collisions on the velocity dispersion of
the clouds is not known.  The internal structure, i.e., magnetic
fields, will play a huge role in determining the efficiency of
cloud-cloud collisional driven viscosity (see for instance Krolik \&
Begelman 1988).  Another way by which this final ``10 pc problem'' can
be solved is via a disk which is supported either by radiation
pressure (Thompson et al.  2005), supernova driven turbulence (Wada \&
Norman 2002; Kawakatu \& Wada 2008), or cosmic ray pressure (Socrates,
Davis, \& Ramirez-Ruiz 2008).  The gas disk or ring which results from
the deposition of gas at ten to tens of parsecs may form a marginally
stable disk which is supported by star formation/supernova which can
then viscously accrete.  One-zone calculations by Thompson et al.
(2005) suggest that a star formation rate of $\sim 1\,{\rm
  M}_{\odot}\,{\rm yr}^{-1}$ is sufficient to support a marginally
stable disk at ten to tens of parsecs (see Thompson et al.  2005,
their Figure 5),\footnote{I note, however, that they presume some sort
  of global torque via spiral arms or the like in their calculations.}
while also allows for a nontrivial accretion rate onto the central
black hole.  I note that the strengths of the ``bars within bars'' or
gas shepherding scenarios and that of the starburst supported disk
scenario are complementary to one another.  Inside of tens of parsecs,
modest star formation rates can support a radiation pressure supported
disk rather than requiring $200-600\,{\rm M}_{\odot}\,{\rm yr}^{-1}$
star formation rates from 200 pcs inward (again see Thompson et al.
2005, their Figure 5).  Recent observation of NGC 3227 by Davies et
al. (2006) support this notion of star formation in a gas disk at
small radii.  However, the current star formation rate of $\sim 0.1\
{\rm M_{\odot}}\,{\rm yr}^{-1}$ appears to be insufficient to support
the vertical scale of the disk.  Other forms of pressure support,
i.e., cosmic ray pressure (see Socrates, Davis, \& Ramirez-Ruiz 2007
their appendix E), may be important in this context.

\section{Discussion and Conclusions}\label{sec:conclusions}

I have calculated the action of an infalling satellite on a coplanar
gaseous disk.  The satellite falls inward because of dynamical
friction and couples to the gaseous disk via disk tides.  Due to the
unrelenting loss of angular momentum, the satellite-disk system
shrinks in radius.  Thus, gas is transported from larger radii to
smaller radii.  I calculate the structure of the evolving disk as a
function of radius, both numerically and analytically and show a
significant density enhancement at the disk edge.  This density
enhancement is due to the transport of gas initially at larger radii
to a ring that is $r_{\rm H,d}$ thick radially.

The fate of this shepherded gas is not known.  I presented two
possible outcomes: 1. the gas will fragment and form stars or 2. the
gas is shepherded into the nuclear region.  In the latter case, I
suggest that the gas shepherding scenario, which I have presented may
be one means by which the ``100 parsec'' problem may be resolved.  For
this to occur, gas must be shepherded sufficiently rapidly such that
the gas is not completely consumed in star formation, which requires a
sufficiently massive satellite ($\sim 10^6\,{\rm M}_{\odot}$).  I
argue that such massive satellites exist on the form of SSCs.  

In the model presented, the physics of star formation is a disk
environment has not been fully addressed.  Observationally, galactic
star-forming disks appear to be marginally unstable to fragmentation,
i.e., $Q \sim 1$ (Thompson et al 2005) and their star formation rate
follows the Kennicutt law (Kennicutt 1998).  These two points suggests
that some sort of feedback is responsible for maintaining this careful
balance between fragmentation and heating.  How this feedback operates
is not well understood.  As discussed previously, various models have
invoked radiation pressure (Thompson et al. 2005), supernova feedback
(Wada \& Norman 2002), and cosmic ray pressure (Socrates et al. 2008)
from star formation.  Alternatively, the turbulent pressure support
may result from gravitational instability, i.e., gravitoturbulence
(Gammie 2001).  It is unclear that in the presence of gas shepherding
if this equilibrium can be maintained at the edge of the disk where
the surface density is enhanced.  If it does not, this edge is likely
to fragment completely into stars on a few dynamical times.  It is
also unclear if the Kennicutt law would continue to hold at the disk
edge.  Additional insights from future research on self-gravitating
star-forming galactic disks will be needed to help formulate a more
self-consistent calculation of the gas shepherding scenario, which
will be needed to fully address the fate of the shepherded gas.

Assuming that the gas is not fully consumed in star formation, it
would be shepherd down to 10 to 10s of parsecs (see \S \ref{sec:final
  10 pc}). At that radii cloud-cloud collisions (Krolik \& Begelman
1988; SBF90) or a starburst supported disk (Thompson et al.  2005)
would allow the material to be viscously accreted.

%The region where the gas would be
%shepherded down into does strongly depend on the central density of
%the SSC.  However, 10 to 10s of parsecs does not appear to be
%unreasonable.

The initial satellite inclination and/or eccentricity is likely to
significantly affect gas shepherding.  In this paper, I have studied a
special case where the orbit of the satellite is in the plane of the
disk and approximately circular for simplicity.  An eccentric
satellite would have its eccentricity damped via the excitation of
spiral density waves.  If the eccentricity is below $r_{\rm H,
  d}/r_{\rm d}$, i.e., the satellite stays within one ``Hill radius''
of the disk of its outer edge, then the shepherding
calculation presented should be a reasonable estimate of the
shepherding timescales and velocities.

Compared to a coplanar satellite, an inclined satellite would couple
less well to the gaseous disk.  For a sufficiently large inclination,
the gas disk and satellite would be effectively decoupled.  However,
its inclination could be sufficiently rapidly damped via induced
bending waves in the gas disk in addition to spiral density waves (see
for instance Ostriker 1994; Terquem 1998).  Once its inclination is
damped below the critical value, where the disk and satellite become
well coupled, the shepherding scenario presented in this paper would
likely follow.  I would expect this critical inclination to be of
order $\sin i = r_{\rm H,d}/r_{\rm d}$, i.e., the satellite stays
within of order a ``Hill radius'' of the disk above and below it.

Extending the present calculation to satellites of modest inclination
would be fruitful and would allow a study of using the nucleated cores
from minor mergers as shepherding satellites (see \S\ref{sec:shepherd
  destruction}). As discussed in \S\ref{sec:applications}, minor
mergers may drive nuclear activity.  The initial inclination of minor
mergers is arbitary, but their inclination is damped as they interact
with the {\it large scale} stellar and gaseous disk.  As a result,
they may approach the shepherding scenario presented in this paper.
Such a study is a subject for future work.

In addition, dynamical friction would also act on the
satellite-induced spiral density waves.  Dynamical friction of the
extended mass perturbation, i.e., the satellite-induced spiral density
waves, may enhance the transfer of angular momentum between the gas
disk and background stars (Tremaine \& Weinberg 1984).  A study of
these effects, while fruitful, is beyond the scope of this work.

The physics of gas shepherding is also important in other areas as
well.  For instance, it may shape radial distribution of star clusters
in nuclear gas rings.  In some of these nuclear ring systems, the
location of the star clusters is exterior to the gas ring from which
they are presumably formed.  For instance, NGC 4314 clearly show the
star clusters exterior to the gas (Benedict et al. 2002).  Martini et
al. (2003) found that the eight galaxies with strong bars all have
star formation exterior to the dust ring in their sample of 123
galaxies.  How gas shepherding drives this morphology is the subject
of a forthcoming work. (Van der Ven \& Chang, in preparation).

%Gas shepherding may help tranport gas between the ILR of the large-scale 
%galactic bar and the accretion disk which fuels the central
%starbursts and AGNs.  In addition, gas shepherding may be a mechanism
%which may operate in early starburst galaxies and hence drive the
%formation of the first black holes (BHs).  One possibility is that the
%first stars in the first galaxies have such low metallicity that
%stellar winds do not result in much mass loss.  Hence the first
%massive stellar clusters would not be disrupted due to mass loss. As a
%result these clusters may participate in shepherding part of the gas
%disk in which they where initially formed into smaller radii where it
%fuels a nascent black hole.  Thus, gas shepherding may play a
%significant role in building the first BHs.

%Nonlocal damping of spiral density waves may result in substantial
%accretion in the inner disk.  Spiral density waves may either be
%damped by viscous processs (Goldreich \& Tremaine 1978, 1980) or wave
%steepening (Goodman \& Rafikov 2001).  Typically, the length scale for
%both these processes is of order the vertical scale height of the
%disk.  As I have chosen a small disk mass and $Q\sim 1$, the vertical
%scale height is extremely small.  As a result, damping of the
%generated spiral density waves are likely to be local.  

\acknowledgements

I thank N. Murray for his early encouragement and useful discussions.
I thank A. Socrates for his advice and continuing encouragement.  I
thank R.  Levine and A. Kratsov for making a early copy of their
manuscript available.  I thank E. Quataert for useful discussions and
a careful reading of this manuscript.  I also thank G. Van der Ven and
E. Chiang for useful discussions.  I thank M. Jones for a careful
reading of this manuscript.  I thank the anonymous referees for very
useful comments which improved the presentation of this paper.  I also
thank the Kavli Institute for Theoretical Physics and the Max-Planck
Institute for Astrophysics for their hospitality during the initial
and final stages of this project, respectively.  I gratefully
acknowledge the support of the Miller Institute for Basic Research.
This research was supported in part by the National Science Foundation
under Grant No. PHY05-51164

\appendix

\section{Calculation of $\sigma_{\rm max}$}\label{sec:sigma max}

I now discuss the calculation of $\sigma_{\rm max}$ and hence derive
the constant of integration $C$.  I first plug equation (\ref{eq:sigma
  approx1}) into equation (\ref{eq:master approx2}) to find
\begin{equation}
x_{\rm d,0}^{-1}\frac {\partial x_{\rm s}} {\partial t'} = {2\beta'} \left(\frac {x_{\rm s}}{x_{\rm d,0}}\right)^3  \int {\sqrt{ C + \frac {q'} {\alpha' q_{\rm d}^{1/3}}\left(-\frac 2 3 v_{\rm shep} x + \frac {2\beta'}{9}\left(\frac{x_{\rm s}}{x_{\rm d,0}}\right)^2 \frac {q'} {q_{\rm d}^{2/3}} 
  \left(\frac {1} {x^{3}}\right)\right)}}\frac {dx} {x^4}
-\frac {x_{\rm d,0}}{x_{\rm s}}.
\end{equation}
I make the approximation that the torque due to the satellite-disk
interaction and the torque from dynamical friction nearly cancel each
other out, i.e., $x_{\rm d,0}^{-1} \partial x_{\rm s}/\partial t' \ll 1$.
Hence I find,
\begin{equation}\label{eq:integral}
{2\beta'} \int {\sqrt{ C + \frac {q'} {\alpha' q_{\rm d}^{1/3}}\left(-\frac 2 3 v_{\rm shep} x + \frac {2\beta'}{9}\left(\frac{x_{\rm s}}{x_{\rm d,0}}\right)^2 \frac {q'} {q_{\rm d}^{2/3}} 
  \left(\frac {1} {x^{3}}\right)\right)}}\frac {dx} {x^4}
=\left(\frac {x_{\rm d,0}}{x_{\rm s}}\right)^4.
\end{equation}

It is helpful to break the integral in equation (\ref{eq:integral})
into two parts.  I rewrite the integral, $I$, as 
\begin{equation}\label{eq:integral definition}
I \equiv \int \rightarrow
\int_{-\infty}^{x_{\rm peak}} + \int_{x_{\rm peak}}^{x_{\rm max}},
\end{equation}
where $x_{\rm max} < x_{\rm s}$ is defined as where $\sigma_{\rm w}(x > x_{\rm max})
= 0$, i.e., the disk is cutoff by disk tides above this radius, and
consider each integral in turn.  The first integral, $I_1$, is
\begin{equation}
I_1 \equiv \int_{-\infty}^{x_{\rm peak}} {\sqrt{ C + \frac {q'} {\alpha' q_{\rm d}^{1/3}}\left(-\frac 2 3 v_{\rm shep} x + \frac {2\beta'}{9}\left(\frac{x_{\rm s}}{x_{\rm d,0}}\right)^2 \frac {q'} {q_{\rm d}^{2/3}} 
  \left(\frac {1} {x^{3}}\right)\right)}}\frac {dx} {x^4}.
\end{equation}
At $x_{\rm peak}$, $\sigma_{\rm w}=\sigma_{\rm max}$ and it declines as $x$
becomes more negative, i.e., away from the satellite, because of the $v_{\rm shep} x$ term in the square root.
However, the torque density declines even faster, so I may approximate
$\sigma_{\rm w}=\sigma_{\rm max}$ for the purposes of this integral, i.e.,
\begin{equation}
I_1 \approx \int_{-\infty}^{x_{\rm peak}} \frac{\sigma_{\rm max}}
{x^4} dx = \frac 1 3 \frac {\sigma_{\rm max}} {x_{\rm peak}^3}.
\end{equation}

The second integral, $I_2$, is 
\begin{equation}
I_2 \equiv \int_{x_{\rm peak}}^{x_{\rm max}} {\sqrt{ C + \frac {q'} {\alpha' q_{\rm d}^{1/3}}\left(-\frac 2 3 v_{\rm shep} x + \frac {2\beta'}{9}\left(\frac{x_{\rm s}}{x_{\rm d,0}}\right)^2 \frac {q'} {q_{\rm d}^{2/3}} 
  \left(\frac {1} {x^{3}}\right)\right)}}\frac {dx} {x^4}.
\end{equation}
Here, the disk torque increases as $x$ approaches $x_{\rm max}$, but
$\sigma_{\rm w}$ declines due to the $x^{-3}$ term in the square root.
As the linear term $v_{\rm shep} x$ is overwhelmed by $x^{-3}$ as $x$
approaches smaller {\it absolute} values i.e. $x \rightarrow x_{\rm
  max}$, I ignore the linear term and approximate this integral by
\begin{equation}
I_2 \approx \int_{x_{\rm peak}}^{x_{\rm max}} {\sqrt{ C + \frac {q'^2} {\alpha' q_{\rm d}}\frac {2\beta'}{9}  \left(\frac{x_{\rm s}}{x_{\rm d,0}}\right)^2
  \left(\frac {1} {x^{3}}\right)}}\frac {dx} {x^4}.
\end{equation}
Changing variables to 
\begin{eqnarray}
y &=& C + \frac {q'^2} {\alpha' q_{\rm d}}\frac {2\beta'}{9} \left(\frac{x_{\rm s}}{x_{\rm d,0}}\right)^2  
  \left(\frac {1} {x^{3}}\right)\\
dy &=& -\frac {2\beta'}{3} \frac {q'^2} {\alpha' q_{\rm d}}\left(\frac{x_{\rm s}}{x_{\rm d,0}}\right)^2\frac 1 {x^4} dx,
\end{eqnarray}
and performing the integral, I find
\begin{equation}
I_2 \approx \beta'^{-1} \left(\frac {x_{\rm d,0}}{x_{\rm s}}\right)^2\frac {\alpha' q_{\rm d}} {q'^2}
\left[\left( C + \frac {q'} {\alpha' q_{\rm d}^{1/3}}\left(-\frac 2 3 v_{\rm shep} x + \frac {2\beta'}{9} \left(\frac{x_{\rm s}}{x_{\rm d,0}}\right)^2\frac {q'} {q_{\rm d}^{2/3}} 
  \left(\frac {1} {x^{3}}\right)\right)\right)^{3/2}\right]^{x_{\rm peak}}_{x_{\rm max}}.
\end{equation}
At $x_{\rm peak}$ the term in brackets, which I identify as my
approximation for $\sigma_{\rm w}$, is $\sigma_{\rm max}^3$, while at $x_{\rm
max}$, the term in brackets is $0$.  Thus I find
\begin{equation}
I_2 = \beta'^{-1}\left(\frac {x_{\rm d,0}}{x_{\rm s}}\right)^2 \frac {\alpha' q_{\rm d}} {q'^2}\sigma_{\rm max}^3
\end{equation}
Hence equation (\ref{eq:integral definition}) becomes
\begin{equation}
I = I_1 + I_2 \approx \frac 1 3 \frac {\sigma_{\rm max}} {x_{\rm peak}^3}+\beta'^{-1}\left(\frac {x_{\rm d,0}}{x_{\rm s}}\right)^2 \frac {\alpha' q_{\rm d}} {q'^2}\sigma_{\rm max}^3.
\end{equation}
Plugging this result into equation (\ref{eq:integral}) gives a cubic
equation for $\sigma_{\rm max}$
\begin{equation}\label{eq:sigma max cubic app}
\frac {2\beta'}{3} \frac {\sigma_{\rm max}} {x_{\rm peak}^3}+2\left(\frac {x_{\rm d,0}}{x_{\rm s}}\right)^2 \frac {\alpha' q_{\rm d}} {q'^2}\sigma_{\rm max}^3 - \left(\frac {x_{\rm d,0}}{x_{\rm s}}\right)^4= 0,
\end{equation}
which I may solve numerically.

\end{document}